\documentclass[twocolumn, prl, superscriptaddress]{revtex4-1}
\usepackage{}
\usepackage{amssymb}
\usepackage{amsmath}
\usepackage{epsfig}
\usepackage{color}
\usepackage{graphics, graphicx}
\usepackage{bbold}
\usepackage{psfrag}
\usepackage{mathcomp}
\usepackage{subfigure}
\usepackage{verbatim}
\usepackage{color}
\usepackage[colorlinks, citecolor=blue]{hyperref}

\begin{document}

%\title{P-wave scattering and molecules in quasi-1D atomic gases}
\title{Many-body stabilization of a resonant p-wave Fermi gas in one dimension}
\author{Lei Pan}
\affiliation{Beijing National Laboratory for Condensed Matter Physics, Institute of Physics, Chinese Academy of Sciences, Beijing 100190, China}
\affiliation{University of Chinese Academy of Sciences, Beijing 100049, China}
\author{Shu Chen}
\affiliation{Beijing National Laboratory for Condensed Matter Physics, Institute of Physics, Chinese Academy of Sciences, Beijing 100190, China}
\affiliation{University of Chinese Academy of Sciences, Beijing 100049, China}
\affiliation{Collaborative Innovation Center of Quantum Matter, Beijing, China}
\author{Xiaoling Cui}
\email{xlcui@iphy.ac.cn}
\affiliation{Beijing National Laboratory for Condensed Matter Physics, Institute of Physics, Chinese Academy of Sciences, Beijing 100190, China}
\date{\today}
\begin{abstract}
Using the asymptotic Bethe Ansatz, we study the stabilization problem of  the one-dimensional spin-polarized Fermi gas  confined in a hard-wall potential with tunable p-wave scattering length and finite effective range. We find that the interplay of two factors, i.e., the finite interaction range and the hard-wall potential, will stabilize the system near resonance. The stabilization occurs even in the positive scattering length side, where the system undergoes a many-body collapse if any of the factors is absent. At p-wave resonance, the fermion system is found to feature the ``quasi-particle condensation" for any value of effective range, which is stabilized if the range is above twice the mean particle distance. Slightly away from resonance, the correction to the stability condition linearly depends on the inverse scattering length. Finally, a global picture is presented for the energetics and stability properties of fermions from weakly attractive to deep bound state regime. Our results raise the possibility for achieving stable p-wave superfluidity in quasi-1D atomic systems, and meanwhile, shed light on the intriguing s- and p-wave physics in 1D that violate the Bose-Fermi duality.  %Our results suggest a many-body stable p-wave Fermi gas is within the reach of current cold atoms techniques.
\end{abstract}

\maketitle

{\it Introduction.} Experimental realization of p-wave ultracold Fermi gases with tunable interactions\cite{K40,K40_2,Toronto,Li6_1,Li6_2,CIR_p_expe} offers great opportunities for investigating many intriguing p-wave phenomena, such as the rich orbital pairing and the topological superfluids\cite{Gurarie, Yip, Read, Zoller, Kitaev}. %In this system, the interaction is highly tunable through p-wave Feshbach resonance or confinement-induced p-wave resonance in low-D\cite{..}.
%To successfully explore the p-wave effects in realistic atomic gas, a key ingredient is the stability of the system. In 3D, the stability is impeded by the severe inelastic three-body loss near the p-wave Feshbach resonance.
%At this stage, a central issue for the experimental detection of low-energy p-wave physics is the severe atom loss near resonance, which prohibits the many-body equilibration in a reasonably long time scale.
However, the detection of p-wave superfluidity of fermions near a Feshbach resonance has encountered great difficulties, due to the severe atom loss which prohibits the many-body equilibration in a reasonably long time scale.  In this context, the (quasi-)1D geometry emerges as a promising candidate to overcome the atom loss due to few-body collisions\cite{Cui, Gora1}.
%{\color{red}\textbf{In 3D and quasi-2D, the loss mainly comes from the strong inelastic three-body collision, which is associated with the presence of centrifugal barrier for two-body collision and the resulted shallow molecules that are highly weighted at short-range and easily decay to deep molecules.}} While such loss is likely to be suppressed in quasi-1D given the spatially extended molecule near resonance\cite{Cui},
Nevertheless, there can be another source for the loss in 1D, i.e., the many-body collapse due to negative compressibility. This is inferred from the theorem of Bose-Fermi duality \cite{duality}, which maps the wave functions and the energies between p-wave fermions and s-wave bosons with inverse coupling strengths \cite{duality,Bender,Girardeau}. Accordingly, the p-wave fermions with a positive scattering length (where a two-body molecule is supported) can be mapped to attractive bosons, which implies the fermions immediately undergo a many-body collapse once across resonance to the molecule side. To ensure the stability, most of previous studies on the 1D p-wave Fermi gas have been carried out in the negative scattering length side\cite{Imambekov, Hu, Chen}, and there has been rare discussions on many-body physics of 1D fermions with positive scattering length.

In this work, we show that a finite effective range of p-wave interaction can save the 1D fermion system from collapsing in both sides of the resonance. Our work is simply motivated by the fact that the Bose-Fermi duality breaks down in the presence of a finite range, as shown from a simple two-body analysis\cite{Cui2}. In reality, the p-wave range of quasi-1D atomic system is reasonably large\cite{CIR_p_1,CIR_p_2,CIR_p_3,footnote_range}, as it is reduced from the large 3D effective range near p-wave Feshbach resonances\cite{K40_2, Toronto}.  Hence, it is also practically important to consider its effect. %Nevertheless, in drawing above conclusion we have missed an important ingredient,  which is the finite effective range. The Bose-Fermi duality only applies to zero range case but not finite range\cite{Cui2}. In practice, the p-wave Feshbach resonance generally has a large effective range in 3D\cite{K40_2, Toronto}, which results in a large range in quasi-1D through the confinement-induced resonance\cite{CIR_p_1,CIR_p_2,CIR_p_3,footnote_range}.  %In a simple two-body problem, it has been shown that the Bose-Fermi duality breaks down in the presence of a finite range. % the finite-range in the fundamental two-body system\cite{..}.  this work, we show that the many-body collapse can be avoided if considering the range effect in 1D Fermi gas.  This is motivated by the fact that the
%Previously, exact treatments of the 1D p-wave Fermi gas have been carried out with periodic boundary at finite range\cite{Imambekov, Hu}, and with open boundary at zero range\cite{Chen}. All these studies are all limited in the negative scattering length side.
%the finite range effect to the ground state energy and excitations of 1D p-wave Fermi gas with periodic boundary have been pointed out in Refs.\cite{Imambekov, Hu}  previously with periodic boundary in the negative scattering length side\cite{Imambekov, Hu}, where the system is known to be stable. The system has also been exactly treated in the same interaction regime in a hard-wall potential without range\cite{Chen}.
%As it is generally believed that fermions with a positive scattering length are unstable, there have been rare discussions of the many-body physics in this regime\cite{textbook}.

Here, using the asymptotic Bethe Ansatz we study the ground state of 1D fermions across p-wave resonance in both positive and negative sides, with a finite interaction range and in a hard-wall potential. It is found that the interplay of the finite range and the hard-wall potential helps to stabilize the system even in the positive scattering length side, where the system would undergo a many-body collapse if any of the two factors is absent. In particular, at p-wave resonance, the fermion system features the ``quasi-particle condensation" for any value of effective range, where all quasi-momenta condense at a single value solely determined by the range and the density. Such a strongly correlated state is found to be stable against collapsing if the range is above twice the mean particle distance. We further extract the stability condition for fermions slightly away from resonance, and finally present a global picture for the energetics/stability property of the system from weakly attractive to deep bound state regime. These results suggest a many-body stable p-wave Fermi gas is within the reach of current cold atoms techniques.
%the correction to the stability condition linearly depends on the inverse scattering length, which can be extracted from the analysis of Bethe {\color{red}Ansatz} equations. Furthermore, we present a global picture for the energetics and the stability property of the system from weakly attractive to deep bound state regime.

{\it Formalism.} To determine the low-energy physics in the quasi-1D regime, i.e., when $E\ll\omega_{\perp}$ ($\omega_{\perp}$ is the frequency of transverse harmonic confinement), we utilize the following boundary condition for the many-body wave function when a pair of fermions come close to each other\cite{Imambekov, Hu} (we set $\hbar=1$ throughout the paper):
\begin{equation}
\lim_{x\equiv x_j-x_i\rightarrow 0^+} \left( \frac{1}{l} + \partial_x - \xi \partial_x^2 \right) \Psi(x_1,x_2,...x_N)=0. \label{BC}
\end{equation}
Here $\{x_i\}$ ($i=1,...N$) are the coordinates of $N$ spin-polarized fermions;
%{\color{red}$l$ is the reduced 1D scattering length, which is highly tunable near the induced p-wave resonance; $\xi$ is the 1D effective range, which stays at a positive constant near resonance\cite{CIR_p_1,CIR_p_2,CIR_p_3, footnote_range}}.
$l$ and $\xi$ are, respectively, the reduced 1D p-wave scattering length and effective range, and near the 1D resonance, $l$ is highly tunable while $\xi$ stays at a positive constant\cite{CIR_p_1,CIR_p_2,CIR_p_3, footnote_range}.

The many-body wave function takes the form:
\begin{eqnarray}
\Psi(x_1,\cdots,x_N)&=&\sum_{Q}\theta(x_{q_N}-x_{q_{N-1}})\cdots \theta(x_{q_2}-x_{q_1})\nonumber\\
&\times&\varphi(x_{q_1},x_{q_2},...,x_{q_N}),
\label{general wavefunction}
\end{eqnarray}
where $\theta$ is the Heaviside step function, $\varphi(x_{q_1},x_{q_2},...,x_{q_N})$ is the wave function for the region $0\leq x_{q_1}\leq x_{q_2}\leq\cdots\leq x_{q_N}\leq L$ (with $L$ the system length), and $Q=(q_1,...,q_N)$ presents a permutation of the position index of $N$ particles. The antisymmetry of $\Psi$ requires
$\varphi(...x_{i},...x_{j},...)=-\varphi(...x_{j},...x_{i},...)$. Therefore we will only calculate the wave function in the region  $x_{1}<x_{2}...<x_{N}$, i.e., $\varphi(x_{1},x_{2},...,x_{N})$, and the wave function in other regions can be easily deduced according to the antisymmetry requirement. Here we consider the system confined in a hard-wall potential, which satisfies the open boundary condition(OBC)
\begin{eqnarray}
\varphi(0,x_{2},...,x_{N})=\varphi(x_{1},x_{2},...,L)=0
\label{OBC}
\end{eqnarray}
According to the Bethe Ansatz, we expand $\varphi$ by plane-waves:
\begin{eqnarray}
\varphi(x_{1},x_{2},...,x_{N})=\sum_{P,\{r_j\}}\left[A_{P,\{r_j\}} \exp\left(i\sum_{j}r_j k_{p_j}x_{j} \right)\right]. \label{BA wavefunction}
\end{eqnarray}
Here $k_j(>0)$ ($j=1,...N$) presents the quasi-momentum, and $r_j=+1(-1)$ denotes the plane-wave of the $j$-th particle (with coordinate $x_j$) moving from left(right) to right(left) in coordinate space; $P=(p_1,p_2,\cdots,p_N)$ presents a permutation of the momentum index of $N$ particles, and $A_{P,\{r_j\}}\equiv A(k_{p_1},k_{p_2},\cdots,k_{p_N}; r_1,r_2,\cdots, r_N)$ is the superposition coefficients. Substituting Eq.(\ref{BA wavefunction}) into Eq.(\ref{general wavefunction}) and applying the boundary conditions Eqs.(\ref{BC},\ref{OBC}), we arrive at the following Bethe Ansatz equations(BAEs):
\begin{eqnarray}
e^{i2k_jL}&=&\prod_{l\neq j}\frac{i(k_j-k_l)+\left[\frac{\xi}{2}(k_j-k_l)^2+\frac{2}{l}\right]}{i(k_j-k_l)- \left[\frac{\xi}{2}(k_j-k_l)^2+\frac{2}{l}\right]}  \nonumber\\
&&\times\frac{i(k_j+k_l)+ \left[\frac{\xi}{2}(k_j+k_l)^2+\frac{2}{l}\right]}{i(k_j+k_l)-\left[\frac{\xi}{2}(k_j+k_l)^2+\frac{2}{l}\right]}
\label{BAE}
\end{eqnarray}
The eigen-energy of the system follows $E=\sum_{j=1}^{N} k_{j}^2/(2m)$. %{\color{red}Compare with the BAEs of the systems under OBC with zero-range\cite{Gaudin,Batchelor,Chen}, here Eq.\ref{BAE} include an additional term which is proportional to the square of relative momentum $\frac{\xi}{2}(k_j-k_l)^2$.}
Note that the second term in the right side of Eq.\ref{BAE} is due to the reflection of particles at the hard-wall boundary, which is uniquely present for the open boundary but absent for periodic boundary condition (PBC)\cite{Imambekov, Hu,footnote_k}. Because of such term, the open boundary fermions can have distinct properties compared to the periodic ones, as will show in this paper. %({\color{red}The sentence "here Eq.\ref{BAE} include an additional term representing the reflection of particles at the hard-wall boundary" has been rewritten.})
%, here Eq.\ref{BAE} include an additional term representing the reflection of particles at the hard-wall boundary. %We also find an error in the BAEs in Refs.\cite{Imambekov, Hu}, which we have corrected in our case\cite{footnote_k}.

Accordingly, the wavefunction $\varphi\left( x_{1},x_{2},\cdots ,x_{N}\right)$ is
\begin{widetext}
\begin{equation}
\varphi=\sum_P (-1)^PA_P\exp \left[ i\left( \sum_{l<j}^{N-1}\omega
_{p_lp_j}\right) +ik_{p_N}L \right] \sin ( k_{p_1}x_{1}) \prod_{1<j<N}\sin \left( k_{p_j}x_{j}-\sum_{l<j}\omega_{p_lp_j}\right) \sin \left( k_{p_N}(
x_{N}-L) \right)
\end{equation}
with $\omega_{ab}=\arctan \frac{\frac{\xi}{2}(k_a-k_b)^2+\frac{2}{l}}{k_a-k_b}-\arctan
\frac{\frac{\xi}{2}(k_a+k_b)^2+\frac{2}{l}}{k_a+k_b}$,  and $A_{P}\equiv A(k_{p_1},k_{p_2},\cdots,k_{p_N}; r_1=r_2\cdots=r_N=1)={\cal N}\prod_{j<l}^{N}\left(ik_{p_j}-ik_{p_l}+[\frac{\xi}{2}(k_{p_j}-k_{p_l})^2+\frac{2}{l}]\right)$, where ${\cal N}$ is the normalization factor. Here
$(-1)^P=\pm 1$ is the sign factor associated with even/odd
permutations of $P=(p_1,p_2,\cdots,p_N)$.
Apparently $\varphi$ satisfies the OBC (\ref{OBC}).
\end{widetext}

For zero-range case ($\xi=0$), one can see that Eq.\ref{BAE} directly reduces to the BAE of identical bosons with coupling $c=-2/l$\cite{textbook}, hence the two systems have the same quasi-$k$ distribution and thus the same $E$, as is exactly predicted by the Bose-Fermi duality\cite{duality}. However, when we turn on a finite range ($\xi>0$), Eq.\ref{BAE} can no longer be reduced to the BAE of the finite-range bosons, due to distinct forms of the energy-dependent coupling strengths for two systems\cite{Cui2}, and thus the duality breaks down. As a result, the physics we will address below for the finite-range fermions will have no correspondence in bosons with zero or finite range\cite{Gaudin,Batchelor,Gurarie2, Qi}. %In the following, we will show that the presence of a finite range, together with the hard-wall potential, will significantly change the energetics and stability of fermions, which has no correspondence in boson system.
In the rest of the paper, we use $L$ and $E_0=(2mL^2)^{-1}$ as the unit of length and energy respectively.

\begin{figure}[h]
\includegraphics[width=8.5cm]{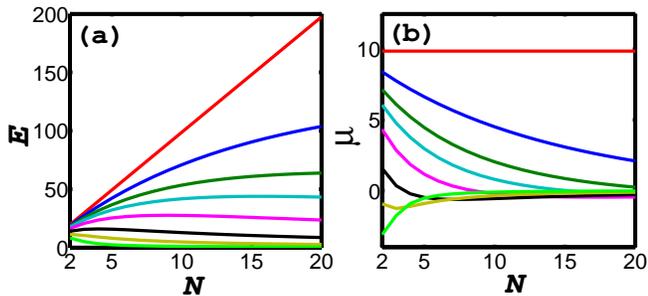}
\caption{(color online) Ground state energy $E$ (a) and chemical potential $\mu$ (b) as functions of $N$ at p-wave resonance. Several different ranges are chosen as (from top to bottom): $\xi=0,\ 0.02,\ 0.04,\ 0.06,\ 0.1,\ 0.2,\ 0.4,\ 1.0$. The units of length and energy are $L$ and $E_0$, respectively.} \label{fig_res}
\end{figure}

{\it At resonance.} We first analyze the range effect at resonance ($1/l=0$). In this case, we find the BAEs (\ref{BAE}) support the ground state with all equal $k_j\equiv k$, which lead to a single closed equation for $k$:
\begin{equation}
k=\pi-(N-1)\arctan(\xi k). \label {BAE_res}
\end{equation}
Accordingly, the wave function simply reduces to
\begin{equation}
\varphi\left( x_{1},x_{2},\cdots, x_{N}\right)\sim \prod_{j=1}^N \sin\left[ kx_{j}-(j-1)\frac{k-\pi}{N-1}  \right]. \label{wf_res}
\end{equation}
In the case of $\xi=0$, we have $k=\pi$ and $\varphi\sim \prod_{j} \sin(\pi x_{j})$, which means all quasi-particles condense at the lowest single-particle state with zero-point energy $E=N\pi^2$. This can be easily understood in the framework of Bose-Fermi duality, namely, the bosons condense at the lowest energy state in the non-interacting limit.
%from the Bose condensation in the non-interacting limit according to the Bose-Fermi duality.
As increasing $\xi$ from $0$, the Bose-Fermi duality breaks down, while, remarkably,  the picture of ``quasi-particle condensation" still holds true in that all the quasi-$k$ of fermions change synchronously from $\pi$.  Accordingly, in the wave function (\ref{wf_res}) each quasi-particle catches a different phase shift from left to right in the coordinate space, in order to match the OBC (\ref{OBC}). These interesting properties uniquely reflect the finite range effect to 1D resonant fermions, which, as far as we know, has not been discovered in any fermion system before.

For large $N$, Eq.(\ref{BAE_res}) gives the solution $k=\pi/\big[1+(N-1)\xi\big]$, and the total energy is
\begin{equation}
E=\frac{N\pi^2}{\Big[1+(N-1)\xi\Big]^2}. \label{E}
\end{equation}
Therefore $E$ will decrease as increasing $\xi$. This can be understood from the energy-dependent scattering length $l(k)$, which follows $l^{-1}(k)=l^{-1}+\xi k^2$. Because of the zero-point energy in a hard-wall potential, the lowest collision energy ($\sim k^2$) are generally positive, which gives a stronger effective interaction [i.e., a larger $l^{-1}(k)$] as increasing $\xi$, and results in a lower energy $E$. This is to be contrast with the PBC case where the zero-point energy is absent, and we have checked that in this case the fermion energy always stays at zero regardless of $\xi$\cite{footnote}.
%range enhance interaction: paper by Jason and me, Hui and Qiran...

Importantly, the energy expression (\ref{E}) also suggests a tunable stability by $\xi$. From (\ref{E}), one can easily derive the chemical potential $\mu=\partial E/\partial N$ and the inverse compressibility $\chi^{-1}=\partial^2 E/\partial N^2$ as:
\begin{eqnarray}
\mu&=& \frac{\pi^2\big[1-(N+1)\xi\big]}{\Big[1+(N-1)\xi\Big]^3} ; \\
\chi^{-1} &=& \frac{2\pi^2\xi\big[(N+2)\xi-2\big]}{\Big[1+(N-1)\xi\Big]^4} .
\end{eqnarray}
The system is stable when $\chi>0$, which, thus, occurs when
\begin{equation}
%\xi>\xi_{p,c}=2d,\ \ \ \ \Big(d=\frac{1}{N+2}\Big)
\xi>\xi_{c}=\frac{2}{N+2}. \label{xi_c}
\end{equation}
In large large $N$ limit, $\xi_{c}\rightarrow 2d$ where $d=1/N$ is the mean inter-particle distance (recalling that the length unit is the system size $L$). This means that with an open boundary, the system can be tuned to be stable once the effective range is above twice the inter-particle distance. Again, this is very different from the PBC case\cite{Imambekov, Hu} or zero-range case\cite{Chen}, where the system is only stable for negative scattering lengths and the stability is not tunable by effective range. Hence, our system is stabilized by the interplay of two essential factors, namely, the presences of a finite range and an external (hard-wall) confinement.

\begin{figure}[b]
\includegraphics[width=9cm,height=7.8cm]{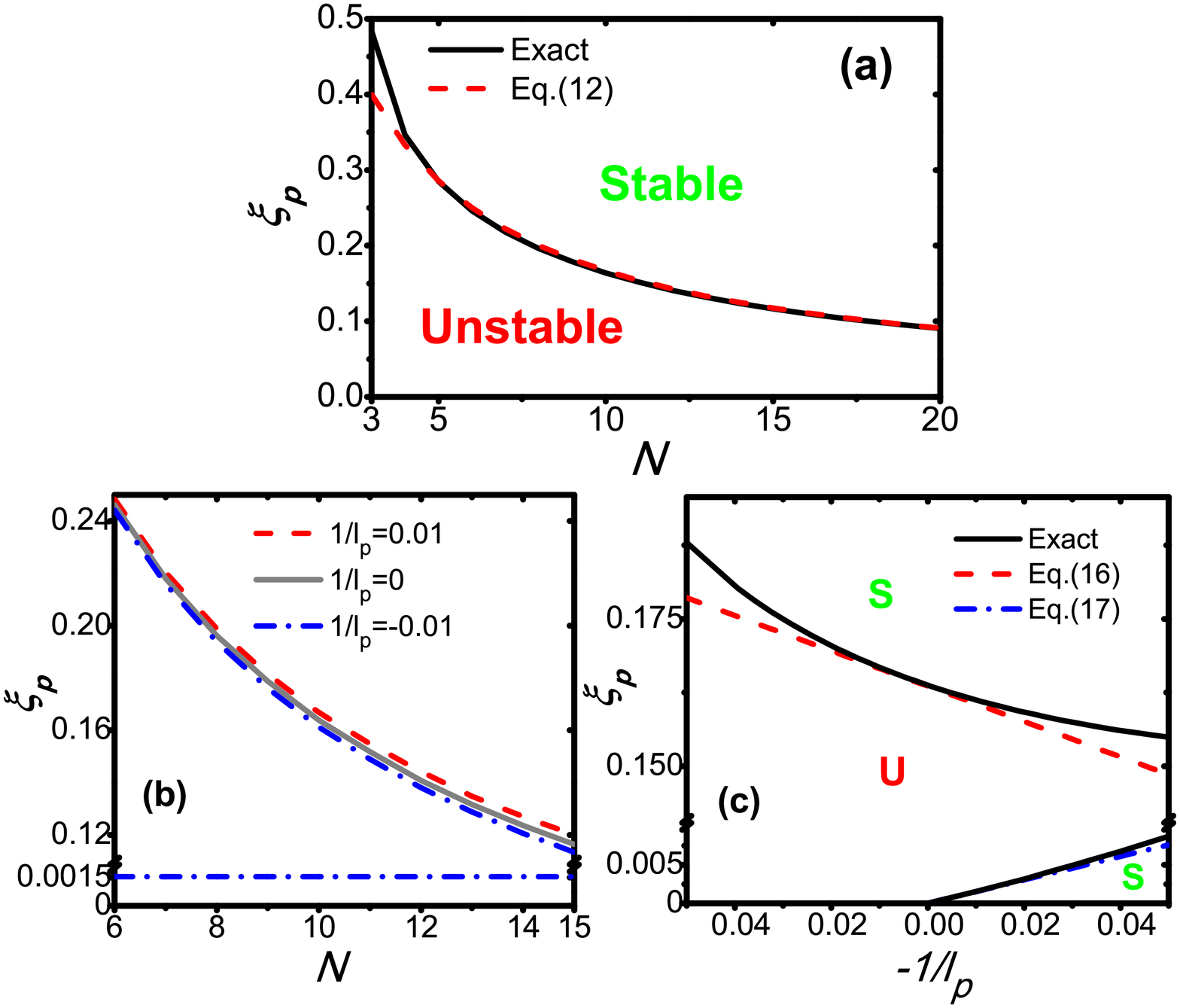}
\caption{(color online) (a) Critical effective range $\xi_{c}$ as functions of $N$ at resonance. The system is stable (unstable) when $\xi$ is above (below) $\xi_{c}$. Red dashed line shows the analytical fit according to Eq.(\ref{xi_c}). (b) $\xi_{c}$ as functions of $N$ near resonance at $1/l=0.01$ (red dashed) and at $1/l=-0.01$ (blue dash-dot). For comparison, $\xi_{c}$ at resonance are shown as gray line. (c) $\xi_{c}$ as changing $1/l$ for $N=10$. "S" and "U" denote respectively the stable and unstable regions. The red dashed and blue dash-dot lines show linear fits to Eq.(\ref{stability_1}) and Eq.(\ref{stability_2}) respectively.
% the resonant critical range $\xi_{p,c}(\infty)$ as changing $1/l$ near resonance for $N=10$. Red solid line is the numerical solution of Bethe ansatz equations. Blue dashed line shows analytical fit based on Eq.\ref{stability}.
} \label{fig_phase}
\end{figure}

%Above analytical results have been confirmed by our numerical calculations.
In Fig.\ref{fig_res}(a,b), we plot $E(N)$ and $\mu(N)=E(N+1)-E(N)$ of the ground state by solving Eq.(\ref{BAE_res}) for different values of $\xi$. We see that as $\xi$ increases from zero, $E$ gradually decreases from the zero-point energy $N\pi^2$, and even changes the curvature as a function of $N$. Accordingly, $\mu$ changes from a constant $\pi^2$ to a varying function of $N$. When $\xi$ is above a critical value $\xi_{c}$, the slope of $\mu(N)$ changes from negative to positive, implying the system become stable with a positive compressibility. In Fig.\ref{fig_phase}(a), we show the numerically extracted $\xi_{c}$ as a function of $N$, which fits well with the analytical prediction (\ref{xi_c}) for large $N(\ge 5)$.

{\it Near resonance.} We now turn to the near resonance regime, i.e., $1/l\rightarrow 0^{\pm}$, where we will extract corrections to the energy and the stability condition up to the lowest order of $1/l$. Away from resonance, the quasi-momenta $k_j$ are no longer identical but depend on $j$. In the regime $1/l\rightarrow 0^-$, all $k_j$ are real and we find that to the leading and next-leading orders it can be expanded as $k_j=k+c_j|l|^{-1/2}+d_j|l|^{-1}$; here $k$ is the solution at $1/l=0$ following Eq.(\ref{BAE_res}). By expanding the BAEs (\ref{BAE}) up to the order of $1/l$, we obtain the following equations for $\{c_j\}$ and $\{d_j\}$:
\begin{eqnarray}
c_j&=&\sum_{l\neq j}\left( \frac{2}{c_j-c_l}-\frac{\xi}{2}(c_j-c_l)-\frac{\xi}{2}\frac{c_j+c_l}{1+(k\xi)^2}\right) ; \label{cj} \\
d_j&=&\sum_{l\neq j}\left( F_{jl} + G_{jl} +\frac{1}{1+(k\xi)^2}\big[\frac{1}{k}-\frac{\xi}{2}(d_j+d_l)\big] \right); \label{dj}
\end{eqnarray}
with $F_{jl}=-(d_j-d_l)(\xi/2+2(c_j-c_l)^{-2})$, $G_{jl}=(c_j+c_l)^2k\xi^3/\big[4(1+k\xi^2)^2\big]$. In the regime $1/l\rightarrow 0^+$, $k_j$ has an imaginary part scaling as $l^{-1/2}$ and we find it can be expanded as $k_j=k+ic_j l^{-1/2}-d_j l^{-1}$, where $\{c_j\}$ and $\{d_j\}$ satisfy the same sets of equations as (\ref{cj},\ref{dj}). Therefore the energy correction up to the order of $1/l$ has a unified form for different sides of resonance, i.e., $\Delta E=(-1/l)\sum_j (c_j^2+2kd_j)$.

In the case of $\xi=0$, we have $c_j=\sum_{l\neq j} \frac{2}{c_j-c_l}$ and $d_j=(N-1)/\pi$, thus the energy correction $\Delta E=(-1/l)3N(N-1)$ is identical to the interaction energy of weakly interacting bosons\cite{Batchelor}, consistent with the Bose-Fermi duality. When $\xi\neq 0$, the situation can be simplified in large $N$ limit, where $k\xi\ll1$ and $F_{jl}$, $G_{jl}$ are negligible compared to the last term in Eq.(\ref{dj}). In this case, we have $\big[1+(N-1)\xi\big]c_j=\sum_{l\neq j} \frac{2}{c_j-c_l}$ and $d_j=(N-1)/\big(k[1+(N-1)\xi]\big)$, and the energy correction is
\begin{equation}
\Delta E=-\frac{1}{l} \frac{3N(N-1)}{\big[1+(N-1)\xi\big]}. \label{dE}
\end{equation}
Given $\Delta E$, we obtain the stability condition near resonance:
\begin{equation}
%\xi\in \left\{\begin{array}{l}(2d+9N^2/\big[2(N+2)^2\pi^2l\big],\infty),\ \ {\rm if} \ \ l\rightarrow +\infty; \\ (0, 3/(2\pi^2|l|))\bigcup (2d-9N^2/\big[2(N+2)^2\pi^2l\big],\infty), \ {\rm if} \ \ l\rightarrow -\infty. \end{array}\right. \label{stability}
\xi>\xi_{c}^{\rm res}+\frac{9N^2}{2\pi^2(N+2)^2} \frac{1}{l};  \label{stability_1}
\end{equation}
here $\xi_{c}^{\rm res}$ is the critical value at resonance. In addition, for $1/l\rightarrow 0^-$ the system has an extra stable region:
\begin{equation}
\xi<-\frac{3}{2\pi^2} \frac{1}{l}. \label{stability_2}
\end{equation}

These results suggest that even in the positive side of resonance, the system can still be stable against collapse in a sizable region of $\xi$. In Fig.\ref{fig_phase}(b), we plot the critical $\xi_{c}$ as function of $N$ for small $1/l$ at different sides of resonance, and we find $\xi_{c}$ indeed only slightly deviate from the resonance value $\xi_{c}^{\rm res}$ (the gray line). In Fig.\ref{fig_phase}(c), we show  $\xi_{c}$ for a given $N=10$ as changing $1/l$, and the numerical results fit well with analytical predictions in Eqs.(\ref{stability_1},\ref{stability_2}) near resonance.
% fit well with our numerical calculations near resonance at both sides.
%We remark that such stability property is very different from the PBC case or the zero-range case, where the system immediately becomes unstable (with $\chi<0)$ once across resonance to the positive side. Therefore the stabilization is facilitated by the interplay of both factors, namely, the finite range and the external (hard-wall) confinement.

\begin{figure}[h]
\includegraphics[width=7.5cm]{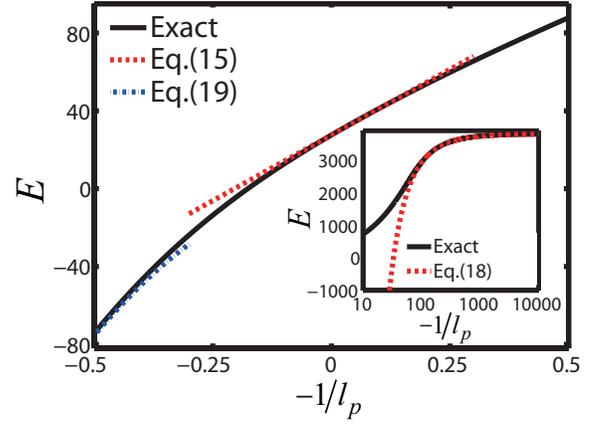}
\caption{(Color online) Ground state energy $E$ as a function of interaction strength $-1/l$ for $\xi=0.1$ at $N=10$. The red dashed line shows the linear fit near resonance based on Eq.(\ref{dE}), and the blue dash-dot line shows the generalized string solution (\ref{string}).  Inset: $E$ in the extremely weak coupling regime, fitted by the perturbation result (\ref{perturb}) (see red dashed).} \label{fig_E}
\end{figure}

{\it Far from resonance.} Further departing from resonance, above results will become invalid when the next-order correction to the energy ($\sim N^3/l^2$) dominates over the lowest one ($\sim N^2/l$) roughly at $|l|< N$. In the weak coupling limit $l\rightarrow 0^-$, the BAEs (\ref{BAE}) give the linear expansion of $k_j$ around the non-interacting value: $k_j=j\pi(1+(N-1)l)$, and this leads to
\begin{equation}
E=E^{(0)}\Big[1+2(N-1)l+o(l^2)\Big], \label{perturb}
\end{equation}
where $E^{(0)}=\pi^2N(N+1)(2N+1)/6$ is the non-interacting Fermi sea energy.
Indeed, such an energy expansion (\ref{perturb}) can also be obtained through the first-order perturbation theory based on the pseudo-potential $U=\sum_{i,j}(2l/m)\partial_{x_{ij}}\delta(x_{ij})\partial_{x_{ij}}$ and an unperturbed Fermi sea. In this limit, the effective range plays negligible role and the system is always stable.

In the deep bound state limit  $l\rightarrow 0^+$, we generalize the $N$-string solution of attractive bosons\cite{Chen,textbook} to the present system as $k_j=\alpha+i(N+1-2j)\sqrt{E_{2b}}$, where $E_{2b}=(1+2\xi/l-\sqrt{1+4\xi/l})/(2\xi^2)$ is the two-body binding energy. For large $N$, the total energy is mainly contributed from the imaginary part of the string solution and is given by
\begin{equation}
E=-\frac{N(N^2-1)}{3}E_{2b}, \label{string}
\end{equation}
which represents a cluster bound state. In this limit, the system is always unstable for any values of $\xi$.

In Fig.\ref{fig_E}, we plot $E$ as function of $-1/l$ for a finite range $\xi=0.1$ at $N=10$. We can see that $E$ increases all along with $-1/l$, and the exact solutions can be well fitted by analytical results  in the weak-coupling limit (\ref{perturb}), near resonance (\ref{dE}), and the cluster bound state limit (\ref{string}). Given all above analyses, we conclude that a finite range will play the most essential role in the resonance regime, where both the energetics (\ref{E},\ref{dE}) and the stability property (\ref{xi_c},\ref{stability_1},\ref{stability_2}) of the system can be significantly modified by the range effect.

%\begin{figure}[h]
%%\includegraphics[width=9cm]{fig4.pdf}
%\caption{Stability diagram in the ($1/l,\xi$) plane at $N=..$} \label{fig_phase2}
%\end{figure}

{\it Discussion and Summary.} Our results can be directly tested in the quasi-1D  Fermi gas with an additional box-trap potential along the (1D) free direction\cite{box_trap}. Take the $^{40}$K Fermi gas near 200G for example, according to the most updated expression  for $\xi$\cite{Cui}, we have $\xi=250 \sim 950$nm for the transverse confinement length $a_{\perp}=60 \sim 120$nm. Therefore $\xi$ can be tuned to stay below or above twice the mean inter-particle distance $d$, which is typically hundreds of nanometers, and thus the stability can be conveniently tuned by $\xi$ or $d$  (according to Eqs.(\ref{xi_c},\ref{stability_1})).
%, and potentially can be used as a guideline for experimentally probing the p-wave effect in a stable many-body system.

In summary, we have shown that a 1D spin-polarized Fermi gas can be stabilized near p-wave resonance (in both sides) in the presence of a finite range and a hard-wall potential, in contrast to the zero-range or periodic boundary condition case where a many-body collapse occurs immediately once across resonance to the positive side. These results can, hopefully, serve as useful guidelines for experiments searching for stable p-wave superfluidity in quasi-1D atomic systems.  Moreover, the range-induced physics revealed in this work, including the phenomenon of quasi-particle condensation and the modified energetics at/near resonance, open up a new avenue of research for 1D boson or fermion systems, in particular, in the regime where the Bose-Fermi duality breaks down and new physics comes out.

% $^6$Li identical fermions, applying a transverse confinement with frequency .., we obtain the 1D effective range $\xi=$ for K and $=..$ for Li, which can be comparable to the typical inter particle distance $d\sim ...$. Further tuning $\omega_{\perp}$ to change $\xi$ or tuning $d$, the stability of the system can be well probed following our result (Fig...).

{\it Acknowledgment.}  The work is supported by the National Key Research and Development Program of China (2018YFA0307601, 2016YFA0300603), and the National Natural Science Foundation of China (No.11622436, No.11425419, No.11421092, No.11534014).


\begin{thebibliography}{99}
\bibitem{footnote_range}Full expressions of the reduced 1D p-wave scattering length and effective range are given in Ref.\cite{Cui}.
\bibitem{footnote_k}
%Ref.\cite{Imambekov, Hu} contain an error in expressing the relative momentum of two colliding fermions (with momenta $k_i,k_j$), which should be $(k_i-k_j)/2$ rather than $k_i-k_j$.
Note that the first terms in the right side of our BAEs (Eq.(\ref{BAE}) in main text) are slightly different from those in Ref.\cite{Imambekov,Hu}. The discrepancy may be induced by their improper use of relative momentum of two colliding fermions (with momenta $k_i,k_j$), which should be $(k_i-k_j)/2$ rather than $k_i-k_j$.
Here, we consider that the correct definition of relative momentum of two colliding fermions is $i(k_j-k_l)/2\sim \partial_x$, which gives rise to Eq.(5) in our paper. However, the BAEs in Ref.\cite{Imambekov, Hu} are consistent with ours if the transformation $\xi\rightarrow\xi/2$, $1/l\rightarrow 2/l$ is applied. %In other words, it's just only a difference of the definition of parameters.
\bibitem{footnote} {In the thermodynamic limit ($N,L\rightarrow\infty$ with $N/L$  fixed), $E$ in Eq.\ref{E} is negligibly small ($\sim 1/N$), which can be seen as the surface energy due to hard-wall potential. Given the zero (bulk) energy in PBC case, here an infinitesimal surface energy can have significant effect to the stability property of the system, as revealed by Eq.\ref{xi_c}.}
\end{thebibliography}
\end{document}